# 'Intelligence Studies Network': A human-curated database for indexing resources with open-source tools

Yusuf A. Ozkan[1,2]

[1] *Department of War Studies, King's College London, London, WC2R 2LS, UK*

[2] *Library Services, Imperial College London, London, SW7 2AZ, UK*

[1] yusuf.a.ozkan@kcl.ac.uk, [2] y.ozkan@imperial.ac.uk - https://orcid.org/0000-0002-3098-275X

## Table of Contents

# Abstract

The Intelligence Studies Network is a comprehensive resource database for publications, events, conferences, and calls for papers in the field of intelligence studies. It offers a novel solution for monitoring, indexing, and visualising resources. Sources are automatically monitored and added to a manually curated database, ensuring the relevance of items to intelligence studies. Curated outputs are stored in a group library on Zotero, an open-source reference management tool. The metadata of items in Zotero is enriched with OpenAlex, an open access bibliographic database. Finally, outputs are listed and visualised on a Streamlit app, an open-source Python framework for building apps. This paper aims to explain the Intelligence Studies Network database and provide a detailed guide on data sources and the workflow. This study demonstrates that it is possible to create a specialised academic database by using open source tools.

# 1. Introduction

The growth of digital databases allows researchers to find publications. However, specific challenges are ahead for intelligence studies. First, the term 'intelligence' is used in different subjects in addition to intelligence studies. As traditional databases cover all academic disciplines, finding sources on intelligence studies may not be straightforward. The results of the search term might be distorted with publications in different fields. To prevent this, the Intelligence Studies Network database is manually curated to make sure that only sources relevant to the field are included. Also, traditional databases include specific types of outputs, predominantly journal articles, conference proceedings or book chapters. Therefore, other types of outputs are systematically excluded from their catalogue. However, the intelligence studies discipline utilizes various sources such as books, theses, reports, podcasts, and magazine or newspaper articles. Although traditional publications databases categorise publications under various topics or themes, they are usually very broad. This database uniquely offers collections/themes exclusively about the field of intelligence studies, such as intelligence history, counterintelligence, and intelligence analysis (collections are detailed below). The 'Intelligence Studies Network' bibliography offers a solution to these problems by indexing sources exclusively about intelligence studies under relevant collections through a manual curation process.

The 'Intelligence Studies Network' database was created in June 2020 to offer a platform for displaying resources on intelligence studies.[1] Its main aim is to form a streamlined platform for individuals seeking to engage exclusively with intelligence sources, eliminating the need to spend time navigating multiple tools to find relevant items. The resources comprise academic and non-academic publications, events, conferences, call for papers, and even academic programs on intelligence studies. A dynamic digest that demonstrates the latest sources or announcements is also available. In addition to listing sources, the website also has various dashboards/metrics that allow users to visualise the intelligence scholarship. The library is updated regularly with newly published sources or old items that are not yet available.

---

[1] Yusuf A. Ozkan, 'Intelligence Studies Network', Intelligence Studies Network, 1 June 2020, https://intelligence.streamlit.app/.

This paper aims to explain the data sources to keep the library updated, the workflow for the database and the website, and interface of the website where sources are listed and visualised.

## 2. Data sources and data collection methods

*Challenges of automating data collection*

The data collection is not fully automated due to the inherent challenges in sourcing information within the field of intelligence studies. The term 'intelligence' is widely used across various disciplines such as artificial intelligence, business intelligence, and cognitive intelligence. The presence of some interdisciplinary works that combine other abovementioned disciplines with the intelligence studies makes identifying sources even more difficult. Even though some databases offer relevant topics about intelligence studies that may seem suitable to automate the process (such as the 'Intelligence Studies and Analysis in Moden Context' topic on OpenAlex), they cover sources that are very loosely or even not connected to the field of intelligence.[2] For example, the book chapter, 'Cryptic alphabets'[3] is listed under the 'Intelligence Studies and Analysis in Moden Context' topic on OpenAlex.[4] Although the word 'cryptic' suggests a relation to intelligence, the content of the work is related to linguistic. It should be noted that it is extremely difficult and even not expected to list all items under the correct theme or topic with great precision on fully automated databases like OpenAlex. However, using such themes within publications databases to automate the process of indexing research publications about intelligence studies would not suit for purpose.

---

Additionally, publication venues for authors can vary. Despite the availability of flagship journals that publish sources primarily on intelligence, some authors choose to publish intelligence-related papers in other journals due to the interdisciplinary nature of the field. For example, a paper covering a content about Cold War intelligence may be suitable for publishing in the Cold War History journal or the Journal of Intelligence History.[5] This means that considering the flagship journals on intelligence studies (these journals are mentioned above) to automate the process of data collection will not be enough to maximise the coverage of publications.

A similar problem is observed for non-academic publications which can be published in various newspapers, magazines, or blog sites. These publication items, which are also known as 'grey literature', are important in disseminating early views of academics or experts about local or global affairs. Authors can also use these venues to publish a shorter version of their journal articles or book. However, similar to the problem explained above, venues do not exclusively publish on intelligence studies. This creates a problem of finding, monitoring, and indexing sources about intelligence.

Therefore, these points make it difficult to fully automate the process of identifying and tracking sources in intelligence. The manual curation is needed to ensure the correct content is included.

*Data collection process*

Academic sources include many publication types like journal articles, books, book chapters, conference papers, and preprints. As journal articles are the most common publication type, an automatic process of tracking articles published in various journals has been developed. OpenAlex which is 'a free and open catalog of the global research system' is used to identify journal articles.[6] OpenAlex was a deliberate choice to retrieve data as it is free and plays an important place in the research infrastructure. OpenAlex content coverage is larger than some known commercial databases and it performs similar to the others in terms of listing references, capturing open access statuses, and finding author country affiliations.[7] It also has a very generous daily limit for API (Application Programming Interface) calls of 100,000 requests per day.[8] By using OpenAlex API, articles published in the last 90 days are retrieved based on the journal name and keywords which are detailed in the next lines.[9]

---

[5] 'Cold War History', Taylor & Francis, accessed 7 August 2024, https://www.tandfonline.com/journals/fcwh20; 'Journal of Intelligence History', Taylor & Francis, accessed 6 August 2024, https://www.tandfonline.com/journals/rjih20.

[6] *OpenAlex Help Center*. https://help.openalex.org/. Accessed 2 Mar. 2024.

[7] Jack Culbert et al., 'Reference Coverage Analysis of OpenAlex Compared to Web of Science and Scopus' (arXiv, 26 February 2024), https://doi.org/10.48550/arXiv.2401.16359; José Luis Ortega and Lorena Delgado-Quirós, 'The Indexation of Retracted Literature in Seven Principal Scholarly Databases: A Coverage Comparison of Dimensions, OpenAlex, PubMed, Scilit, Scopus, The Lens and Web of Science', *Scientometrics* 129, no. 7 (1 July 2024): 3769–85, https://doi.org/10.1007/s11192-024-05034-y; Juan Pablo Alperin et al., 'An Analysis of the Suitability of OpenAlex for Bibliometric Analyses' (arXiv, 26 April 2024), https://doi.org/10.48550/arXiv.2404.17663.

By using OpenAlex, all articles – regardless of their titles – published in the intelligence flagship journals are included in the database – the Zotero group library (the flagship journals are *Intelligence and National Security*, *International Journal of Intelligence and Counterintelligence*, *Studies in Intelligence*, and *Journal of Intelligence History*). [10] For journals that are not predominantly focused on intelligence but feature titles pertinent to intelligence studies, articles containing specific keywords in their titles are considered for inclusion ('intelligence', 'spy', 'counterintelligence', 'espionage', 'covert', 'signal', 'sigint', 'humint', 'decipher', 'cryptanalysis', 'spying', 'spies'). Once the title is manually confirmed that it is relevant to the field of intelligence, it is included in the database. Book reviews are not added to the Zotero group library as a normal item and thus are not available on the final version of the database and the website. However, they can be found as an attached item under the associated book in the Zotero library.

The main source for books and book chapters come from book reviews but some prominent publishers' websites are also tracked by using the same keywords mentioned in the paragraph above. Unlike journal articles, the search for books is not automated as most of the publishers do not have any API service in place. However, Google Scholar alerts for the abovementioned keywords are used to monitor books.

The rest of the item types in the library are gathered by using various automated and non-automated ways, such as tracking RSS feeds where available, Google Scholar alerts, tracking social media channels that share sources on intelligence studies, and monitoring various databases.

For some certain publication types (mostly journal articles), the website also provides additional data in addition to the usual publication metadata. These additional data include citations and open access information at the moment. The citation and open access information are retrieved from OpenAlex with its API by running the DOIs in the dataset against the DOIs in OpenAlex once a week. The enriched dataset is saved into a GitHub repository as CSV from which the certain

---

[8] OpenAlex, 'API Overview | OpenAlex Technical Documentation', 24 April 2024, https://docs.openalex.org/how-to-use-the-api/api-overview.

[9] OpenAlex. *OpenAlex Technical Documentation*. https://docs.openalex.org/. Accessed 2 Mar. 2024.

[10] The names of flagship journals are: 'Intelligence and National Security', Taylor & Francis, accessed 6 August 2024, https://www.tandfonline.com/journals/fint20; 'International Journal of Intelligence and CounterIntelligence', Taylor & Francis, accessed 6 August 2024, https://www.tandfonline.com/journals/ujic20; 'Studies in Intelligence', accessed 6 August 2024, https://www.cia.gov/resources/csi/studies-in-intelligence/; 'Journal of Intelligence History'; Ortega and Delgado-Quirós, 'The Indexation of Retracted Literature in Seven Principal Scholarly Databases'.

Released: August 2024

data points are displayed on the website.[11] The citation count for each journal article is displayed alongside the sources where available, allowing users to easily access the cited papers on OpenAlex. Similarly, if the open access version of a work is found, its location is hyperlinked in addition to the publication link on the website.

*Available item types in the library*

The 'Intelligence Studies Bibliography' database contains metadata of written, audio, and visual sources on intelligence studies. Different types of written sources are available such as journal articles, conference papers, magazine or newspaper articles, blog posts, reports, or documents. The dominant item type of the library, in terms of the number of publications, is journal article which is also the most common publication type in almost every academic field. The following chart illustrates the distribution of sources by publication type in the database. To account for the significant dominance of journal articles, the data is presented on a logarithmic scale.

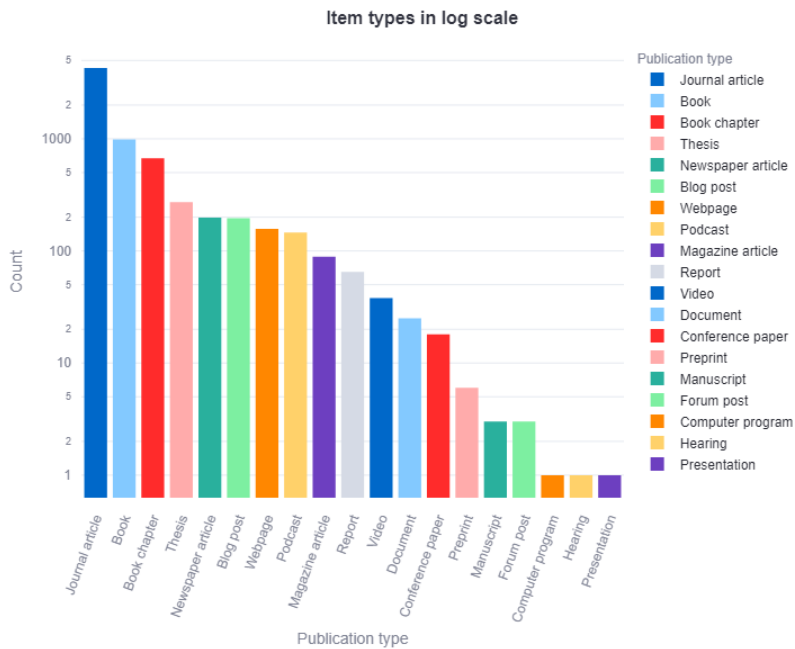

*Figure 1: Available publication types in the database (in logarithmic scale)*

---

Released: August 2024

# 3. Database and website

The main database for publications is Zotero, which is free and open-source reference software.[12] The main reason for this choice is that Zotero is very suitable for keeping different levels of metadata for publications as well as it being open source. Also, Zotero has an option to create shared group libraries which can be managed or used by other people. Even without the Intelligence studies network website on Streamlit, users are still able to use the group library to see and cite sources in the database easily. Zotero also has a plug in for text editing software (such as Microsoft Word or Google Docs) allowing an easy experience for users to cite sources directly. This means that the entire Intelligence Studies Bibliography database is available to use when drafting a research paper with the most common text editor software. It also allows extracting data in the group libraries through its API which is used to retrieve data to the website.[13] Once a source is identified with the methods covered in the previous part, it is included in the shared group library on Zotero, namely '*Intelligence Bibliography*'.[14]

As mentioned above, publication metadata are enriched with citation and open access information via OpenAlex. After the enrichment process, the final dataset is saved into the GitHub repository as CSV.[15] This means that the enriched version of the dataset is not available on Zotero but on the GitHub repository which is also publicly available.[16]

The website is based on Streamlit which is 'an open-source Python library that makes it easy to create and share custom web apps'.[17] Streamlit is well documented and has a well-established developer team and community. Since it allows listing items and visualising the scholarship at the same time, Streamlit was deliberately chosen for this purpose. However, given the database on Zotero and the enriched dataset available on GitHub, the platform for displaying the intelligence studies resources is less important.

---

[12] *Zotero* . https://www.zotero.org/. Accessed 2 Mar. 2024.

[13] *Zotero Documentation: Zotero Web API V3*. https://www.zotero.org/support/dev/web_api/v3/start. Accessed 2 Mar. 2024.

[14] *Zotero Intelligence Bibliography Group Library*. 2020, https://www.zotero.org/groups/2514686/intelligence_bibliography.

[15] Ozkan, Yusuf. "Zotero-Intelligence-Bibliography." Python, February 6, 2024. https://github.com/yusufaliozkan/zotero-intelligence-bibliography.

[16] Ozkan, 'Zotero-Intelligence-Bibliography'.

[17] *Streamlit Documentation*. https://docs.streamlit.io/. Accessed 3 Mar. 2024.

Released: August 2024

## 3.1.     Home page

The home page of the website provides a snapshot of the website and the database in addition to the different search functionalities. Some key metrics are available at the top of the home page. The key metrics include the number of items in the library, total number of citations, average citation (with the option of excluding outliers), open access coverage, number of publication types, total author count, author/publication ratio, and collaboration ratio for the entire database. The last update date of the library is also available.

The home page consists of two tabs: publications and dashboard. The 'Publications' tab is designed to list publications. The first section on the home page after key metrics is 'Search in database' which is dedicated for searching publications within the database with various search options. The 'Search keywords' part allows users to search a keyword or string to list publications whose titles or abstracts contain the searched item. The Boolean operators are available for users in this section. All author names are listed based on the number of publications on the 'Search author' section for users to search author names. It should be considered that this website may not list all publications of the selected author because the paper may not be identified or related to intelligence studies. Users can also list publications by collections, item type, or journal (for journal articles). The library includes items published between 1874 to the present day which can be listed by publication date. The last search option allows users to see journal articles that have been cited at least once. This section has also a 'Trends' module from which you can see the citations occurred in the last two years to the papers published in the same period. This module aims to demonstrate trends in the study of intelligence. From the filters, outlier cited articles (identified as items that have over 1,000 citations) can be excluded from the analysis. All search options have metrics, themes, additional filters, and dashboards displayed on the respective sections.

The Overview section of the home page shows the top 10 recently added and published items. The top 10 cited items can also be found in the same section. This part allows frequent users of the site to keep updated about latest additions or publications. Under the 'All items in database', users can download all items in the database directly from the home page and see the growth of the library over time.

The second tab on the home page is 'Dashboard' where various interactive visuals about the publications in the database are available. The visuals show publications by collection, type, year and author. The dashboard also shows single/multiple authored publications and outputs by open access and citation status. Users can also visualise publishers and journals to find out the main publication venues. A map is available showing country mentions in titles of research outputs. Location, people, and organisation names are retrieved from titles and abstracts with Named Entity Recognition (NER) which is a Natural Language Processing method by using SpaCy – an open-source Python library. The results for NER are displayed in the dashboard section with three charts.

## 3.2.    Collections

The website has sub-pages, each of which list publications belonging to a specific theme in the field of intelligence studies. Each subpage allows users to filter publications by their types and search publications with keywords. Similar to the home page section detailed in the previous section, each individual collection consists of overview, publications, and dashboard sections. The first section covers key metrics about the selected collection, such as the number of items, citation counts, open access coverage, and citation counts. The publications section lists all items within that collection. The last section provides visuals about publications. Some collections contain sub-collections, e.g. the 'Intelligence history' collection has a sub-collection of 'Napoleonic Wars' and 'WW1 (First World War)'.

Currently, the database has 12 main collections[18] each of which are shown in a separate page on the website:

1. **Intelligence history**: This part covers sources about different periods, from pre-Napoleonic Wars to post-Cold War. It also includes sections for 'Terrorism, insurgency, and counterinsurgency' and 'Intelligence archives and methodology'.
2. **Intelligence studies**: The 'Intelligence studies' collection is an umbrella term for many publications related to the discipline, such as 'intelligence and strategy', 'intelligence and culture', 'intelligence education', 'policy and intelligence', and 'media'.
3. **Intelligence analysis:** This section includes sources about intelligence analysis from historical cases as well as its theoretical background.
4. **Intelligence organisations**: This collection includes sources about organisational theories and different intelligence organisations around the world.
5. **Intelligence failures**: Intelligence failure is one of the main research topics in this field. This collection lists publications about intelligence failures, warning, surprise, and politicization of intelligence.

---

[18] Please note that the number of main collections and their names may change quickly as a result of adding, removing, or amalgamating collections. The major changes are announced through the mailing list https://groups.google.com/g/intelligence-studies-network.

6. **Intelligence oversight and ethics**: Resources about accountability, oversight, and ethics on intelligence can be found in this collection.

7. **Intelligence collection**: This collection segregates sources focusing on different methods of intelligence collection, such as HUMINT, SIGINT, OSINT, IMINT etc.

8. **Counterintelligence:** The Counterintelligence collection includes sources from historical cases to the theory of counterintelligence.

9. **Covert action:** This collection covers resources about covert action cases.

10. **Intelligence and cybersphere**: This section lists resources about intelligence and cybersphere. Publications exclusively about cybersphere are not included.

11. **Global intelligence**: The Global intelligence collection shows sources that are about non-UK/US intelligence. This may be the most powerful aspect of the bibliography. Considering that topics related to United Kingdom and United States intelligence history, culture, or organisations have dominated the literature for so long, this collection stands out by shedding light on other nations, particularly those in the 'global south'. Users can easily explore sources from various countries or continents through customizable filters (refer to figures below).

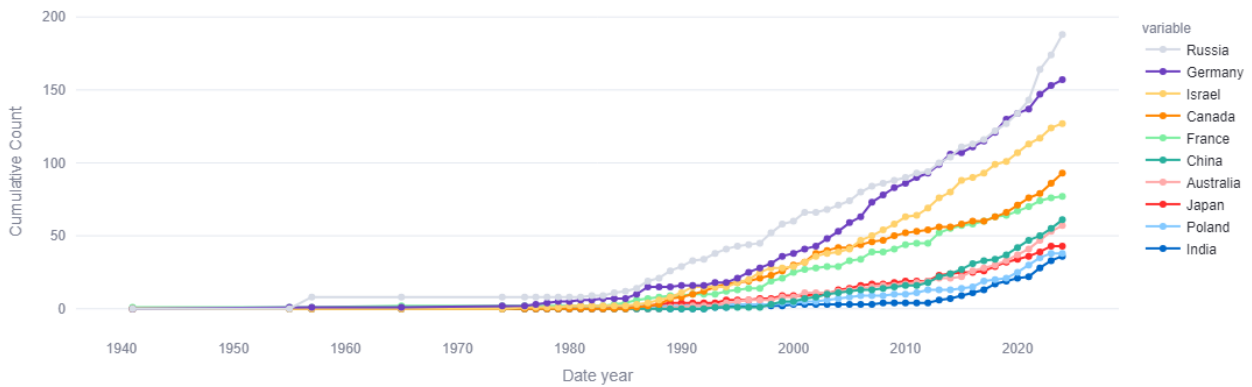

*Figure 2: Cumulative number of publications per country (Top 10 countries)*

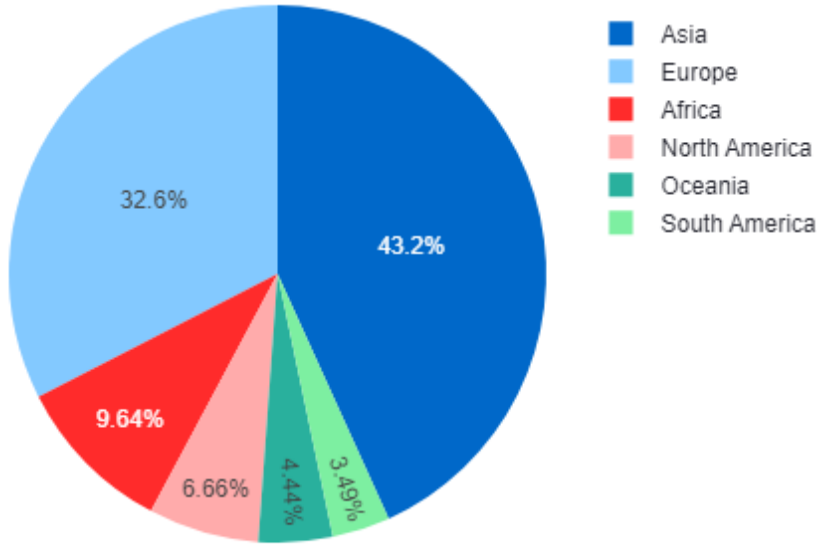

*Figure 3: Number of publications by continent*

12. **Special collections**: This collection is a dedicated space for themes that cannot be categorised under any of the collections mentioned above. It also provides collections for recent conflicts or crises such as the War in Ukraine or the Conflict in the Middle East.

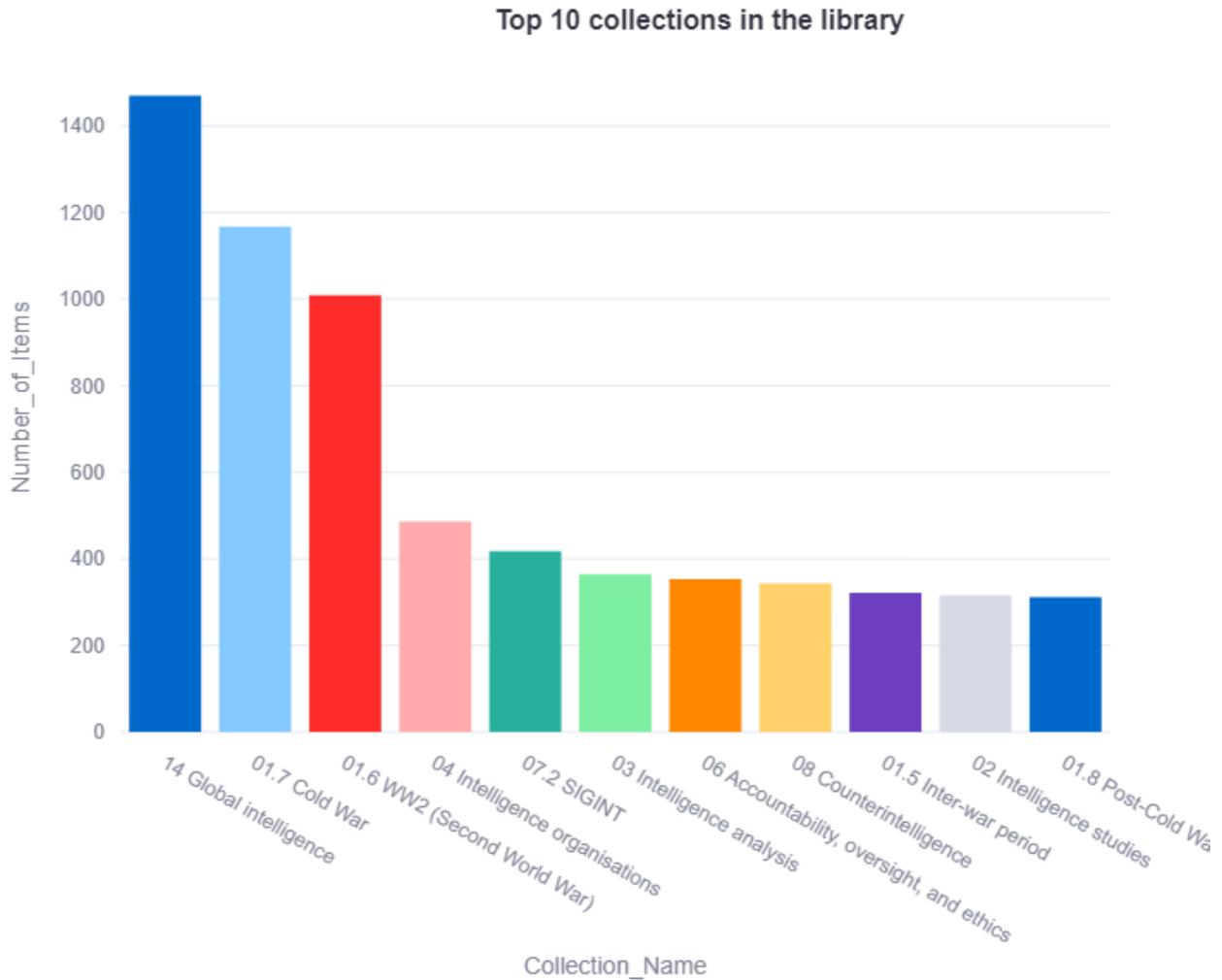

*Figure 4: Top 10 collections in the library*

## 3.3.    Events

This is another section of the website where events, conferences, and call for papers on intelligence studies are listed (https://intelligence.streamlit.app/Events). Events data are not programmatically collected. The data is manually monitored online by tracking major organisations that held events on intelligence. The relevant items are first saved into a database (currently a Google Sheets) and then they are retrieved from the database into the website through its API.[19]

---

[19] Connect Streamlit to a Private Google Sheet - Streamlit Docs. https://docs.streamlit.io/. Accessed 28 Apr. 2024.

Users can sort events by their dates, most recently added to the database, or organizer. The page also displays visuals for the number of events over time and by organizers. Past events are also listed for the benefit of archiving events.

Similarly, conferences and call for papers about the field of intelligence can be found in the same place.

## 3.4.    Digest

The Digest is a section where a snapshot of publications, events, conferences, or call for papers is listed ([https://intelligence.streamlit.app/Digest](https://intelligence.streamlit.app/Digest)). Users can find everything in one place with a more targeted content. This section mostly allows users to see recently added, published, and cited publications with date selection available. It also lists future events, conferences, and call for papers. It is an ideal page for those who check the website regularly to keep informed about the intelligence scholarship and beyond.

## 3.5.    Institutions

The last sub-page of the website is Resources on intelligence studies. This page lists institutions, academic programs, and other resources on intelligence and intelligence studies. This page aims to create a directory for institutions. It currently lists Higher Education Institutions that have an academic program on intelligence studies, government institutions primarily involved in intelligence, museums, research centres focusing of the intelligence field, think tanks, and other resources. However, it is important to note that the list may not be exhaustive as there is no definiteve data source listing all academic or governmental institutions on intelligence.

## 4. Limitations

The Intelligence Studies Network website and database offers the most comprehensive sources on the field of intelligence studies. However, it has some limitations. Firstly, the service is not fully automated. The index is updated manually by the creator. This can bring some advantages such as preventing irrelevant sources from inclusion but also create problems such as missing some sources. Also, since the creator's expertise may not cover all sub-fields of intelligence studies, some sources may not be captured especially if the titles of sources do not sufficiently and clearly signify the relevancy to the field even if the content is relevant.

The Intelligence Studies Network database does not claim to show all outputs of intelligence published in the world. The 'grey literature' items – sources beyond the traditional academic or commercial publishing such as newspaper articles, podcasts, and reports – are included in the database. However, it is not possible to systematically capture and include all grey literature items because most of them need to be manually found online instead of via traditional publication databases. OpenAlex and Google Scholar are used to monitor and include academic sources. Although the two databases are very broad in terms of indexing academic publications, they still may not be able to list some publications. Although sources in other languages are included in the database, their number is quite limited at this phase. This does not mean that scholarly or non-scholarly publications in other languages on intelligence studies are not produced in good quantity.

The database may also not display all publications of the selected author or journal. The author or the journal may publish sources that are not relevant to the field of intelligence studies. These sources are not included in the database. This is especially the case for those working on military history or international relations where interdisciplinary publishing is very common.

Some visuals in dashboard sections should also be used carefully. For example, citation count and open access information may not be found for all items in the database since the two data points are retrieved from OpenAlex which does not index grey literature outputs at all and may not include some academic publications. Similarly, there might be issues with the article metadata – citation count, open access information, and references – available in the OpenAlex database.

Given the limitations mentioned above, this database and the website should never be used to compare research performance of authors, journals or even collections. Its main aim is to index sources and display some key metrics in the field to understand the trends and the development of intelligence studies.

## 5. Conclusion

The Intelligence Studies Network is a human-curated database aiming to index academic and non-academic publications, events, conference, call for papers, and institutions about the field of intelligence studies. It offers a solution to a problem of regularly monitoring, updating, and listing sources on intelligence. Currently, it is one of the most comprehensive databases for publications in terms of source coverage. In addition to listing sources, the website has additional functionalities such as providing metrics and dashboards to visualise the intelligence scholarship.

The website and the database are all based on open-source platforms and have an open licence for everyone to use freely. Publications are recorded in Zotero which is a free and open-source reference management tool. The visual interface is presented through Streamlit which is another open-source Python framework. The publications data are enriched with open access, citations, and references information by using OpenAlex database which is licensed as CC0.

This platform can be used for various purposes. Its main purpose is to find publications on intelligence studies with different search functions and filters. With the available visualisations on dashboards, the website allows for the analysis of trends in scholarly publishing of intelligence changes, including the number of publications per year, citation trends, author-collaboration ratios, and the growth of open access publications.

This study is important because it shows the possibility of creating a database of publications and listing/visualising publications through an external website for specific academic fields by using the combination of different open source tools. The proof of concept explained here can be applied to any academic or non-academic discipline.

# 6. Data and code access statement

No data or software code is utilised for this work. However, the live database for the website is available at the Intelligence Studies Bibliography Zotero Group Library [20] and the zotero-intelligence-bibliography GitHub repository.[21] The source code for the Streamlit website and code to retrieve metadata from OpenAlex are also available in the same repository.

# 7. Conflict of interest

The author declares that no financial or other forms of support were received from any institutions or platforms mentioned in this paper.

---

[20]'Zotero Groups: Intelligence Bibliography', 1 June 2020, https://www.zotero.org/groups/2514686/intelligence_bibliography.
[21] Ozkan, 'Zotero-Intelligence-Bibliography'.

## 8. Licence

'Intelligence Studies Network': A human-curated database for monitoring and indexing resources on intelligence studies © 2024 by Yusuf A. Ozkan is licensed under